\documentclass[fp]{jpsj3}
\usepackage{txfonts}

\usepackage[dvipdfmx]{graphicx}
\usepackage{bm}
\usepackage{color}
\title{Dzyaloshinsky-Moriya Interaction and the Ground State in \textit{S}=3/2 Perfect Kagome Lattice Antiferromagnet $\mathbf{KCr_3(OH)_6(SO_4)_2}$ (Cr-Jarosite) Studied by X-band and High-Frequency ESR}

\author{Susumu Okubo$^{1, 2}$\thanks{sokubo@kobe-u.ac.jp}, Ryohei Nakata$^2$, Shohei Ikeda$^2$, Naoki Takahashi$^2$, Takahiro Sakurai$^3$, Wei-Min Zhang$^1$, Hitoshi Ohta$^{1, 2}$, Tokuro Shimokawa$^4$\thanks{Present address: Department of Earth and Space Science, Graduate School of Science, Osaka University, Toyonaka, Osaka 560-0043, Japan}, T\^{o}ru Sakai$^{5, 6}$, Koji Okuta$^7$, Shigeo Hara$^7$\thanks{Present address: Center for Supports to Research and Education Activities, Kobe University, 1-1 Rokkodai, Nada, Kobe, Hyogo 657-8501, Japan}, and Hirohiko Sato$^7$
}

\inst{
$^1$ Molecular Photoscience Research Center, Kobe University, 1-1 Rokkodai, Nada, Kobe, Hyogo 657-8501, Japan \\
$^2$ Graduate School for Science, Kobe University, 1-1 Rokkodai, Nada, Kobe, Hyogo 657-8501, Japan\\
$^3$ Center for Supports to Research and Education Activities, Kobe University, 1-1 Rokkodai, Nada, Kobe, Hyogo 657-8501, Japan\\
$^4$ Center for Collaborative Research and Technology Development, Kobe University, 1-1 Rokkodai, Nada, Kobe, Hyogo 657-8501, Japan\\
$^5$ Japan Atomic Energy Agency, SPring-8, Sayo, Hyogo 679-5148, Japan\\
$^6$ Graduate School of Material Science, University of Hyogo, Kamigori, Hyogo 678-1297, Japan\\
$^7$ Department of Physics, Chuo University, Bunkyo, Tokyo 112-8551, Japan\\
}

\abst{
A single-crystal \textit{S}=3/2 perfect kagome lattice antiferromagnet, $\mathrm{KCr_3(OH)_6(SO_4)_2}$ (Cr-jarosite), has been studied by X-band and high-frequency electron spin resonance (ESR). 
The \textit{g}-values perpendicular to the kagome plane (\textit{c}-axis) and in the plane were determined to be $g_c=1.9704 \pm 0.0002$ and $g_\xi=1.9720 \pm 0.0003$, respectively, by high-frequency ESR observed at 265 K. 
Antiferromagnetic resonances (AFMRs) with an antiferromagnetic gap of 120 GHz were observed at 1.9 K, which is below $T_\mathrm{N}$=4.5 K. 
The analysis of AFMR modes using the conventional molecular field theory gave $d_\mathrm{p}=0.27$ K and $d_\mathrm{z}=0.07$ K, where $d_\mathrm{p}$ and $d_\mathrm{z}$ are in-plane and out-of-plane components of \textit{d} vector of the Dzyaloshinsky-Moriya (DM) interaction, respectively. 
On the basis of these results and the exchange interaction of \textit{J}=6.15 K estimated by Okuta \textit{et al}., the ground state of Cr-jarosite was discussed in connection with the Monte Carlo simulation results with classical Heisenberg spins on the kagome lattice by Elhajal \textit{et al}. 
Finally, the angular dependence of the linewidth and lineshape observed at 296 K by X-band ESR showed the typical behavior of a two-dimensional Heisenberg antiferromagnet, suggesting the good two-dimensionality of Cr-jarosite. 
}

\begin{document}
\maketitle


%


\section{Introduction}
   In condensed matter physics, frustration has been considered recently as an important and fundamental concept
that plays important roles in a wide range of systems, such as magnets, metals, superconductors, multiferroics, and dielectrics \cite{ref1}. 
Frustration is very interesting because it induces novel effects as a result of enhanced fluctuations, and frustrated systems are expected to fall into an exotic order, a new thermodynamic phase, or novel dynamics. 
To gain deeper insight into frustration, experiments on simple and high-quality frustrated systems 
are required. 
One candidate is a geometrically frustrated magnet. 
Geometrically frustrated magnets, such as triangular, kagome, and pyrochlore antiferromagnets, have been intensively theoretically studied \cite{ref2}. 
Among two-dimensional systems, the kagome antiferromagnet is the most interesting because it is theoretically considered to have the highest frustration, that is, the highest degree of the degenerate state, and is a candidate for the spin-liquid ground state \cite{ref3}. 
However, even the ground state of a kagome antiferromagnet is still under theoretical debate.
For instance, the presence of a gap between the ground state and the first excited state in a kagome antiferromagnet does not yet become clear even by the many numerical calculations \cite{ref4, ref5, ref6}. 
Therefore, experiments on the model substance of the kagome antiferromagnet are strongly desired, but few have been reported \cite{ref7, ref8, ref9, ref10}.
Moreover, the quality of kagome antiferromagnets have exhibited some problems, such as a deficient or deformed lattice or the lack of a single crystal, as in the case, for instance, of $\mathrm{SrCr_{9-x}Ga_{3+x}O_{19}}$ \cite{ref10, ref11}.\\
\indent
   Recently Okuta \textit{et al}. succeeded in synthesizing high-quality single crystals of Cr-jarosite [$\mathrm{KCr_3(OH)_6(SO_4)_2}$] that had no defects in the Cr ions and that were large enough to allow the measurement of anisotropy in magnetization \cite{ref12}. 
Although Cr-jarosite is expected to show magnetic properties intermediate between classical and quantum properties owing to Cr ions (\textit{S}=3/2), it is an almost ideal compound for the study of kagome lattice antiferromagnets because it shows no ion exchange or lattice distortion, which are often seen in other Cu-based or Cr-based kagome lattice model substances \cite{ref7, ref8, ref10, ref11}. 
However, previously reported magnetic properties of Cr-jarosite have been rather controversial, reflecting the difficulty in preparing good powder samples \cite{ref13, ref14, ref15, ref16}. 
$\mu$SR and magnetization results \cite{ref13, ref14} suggested a spin-glass behavior below 2 and 4 K, respectively, while neutron diffraction results suggested a long-range magnetic order with a nearly 
$120^\circ$ structure, although the transition temperature varied among reports \cite{ref15, ref16}. 
Moreover, previous neutron studies suggested a significant deficiency of Cr atoms, that is, 76\% \cite{ref16} and 95\% \cite{ref15} Cr occupancies. 
On the other hand, no evidence of defects in Cr sites was obtained by X-ray diffraction study for a Cr-jarosite single crystal synthesized by Okuta \textit{et al} \cite{ref12}. 
The temperature dependence of magnetic susceptibility that Okuta \textit{et al.} obtained for a high-quality single crystal showed an antiferromagnetic transition accompanied by a weak ferromagnetism at $T_\mathrm{N}$=4.5 K \cite{ref12}. 
The frustration factor $\theta/T_\mathrm{N}$ is estimated to be 14, suggesting strong frustration in the kagome plane of Cr-jarosite. 
The nearest-neighbor exchange interaction \textit{J} was estimated to be 6.15  K following the same analysis using a high-temperature expansion model by Morimoto \textit{et al}. \cite{ref14} and Harris \textit{et al}. \cite{ref17}. 
Moreover, Okuta \textit{et al}. revealed the spontaneous magnetization of 0.05 $\mu_\mathrm{B}$/Cr along the \textit{c}-axis (perpendicular to the kagome plane) below $T_\mathrm{N}$ \cite{ref12}, which became possible by the synthesis of single crystals. 
They discussed the possible origin of weak ferromagnetism being the DM interaction and suggested that the DM vector should be perpendicular to the nearest Cr-Cr bond in the kagome plane on the basis of the crystal symmetry \cite{ref12}. 
However, the determination of a DM vector is rather important because a Monte Carlo simulation with classical Heisenberg spins on the kagome lattice shows that the ground state of a kagome lattice antiferromagnet depends strongly on the parameters $d_\mathrm{p}$, $d_\mathrm{z}$ (in-plane and out-of-plane components of a DM vector), and \textit{J} (the nearest-neighbor exchange interaction) \cite{ref18}. \\
\begin{figure}[tbp]
\begin{center}
\includegraphics[width=5.5cm]{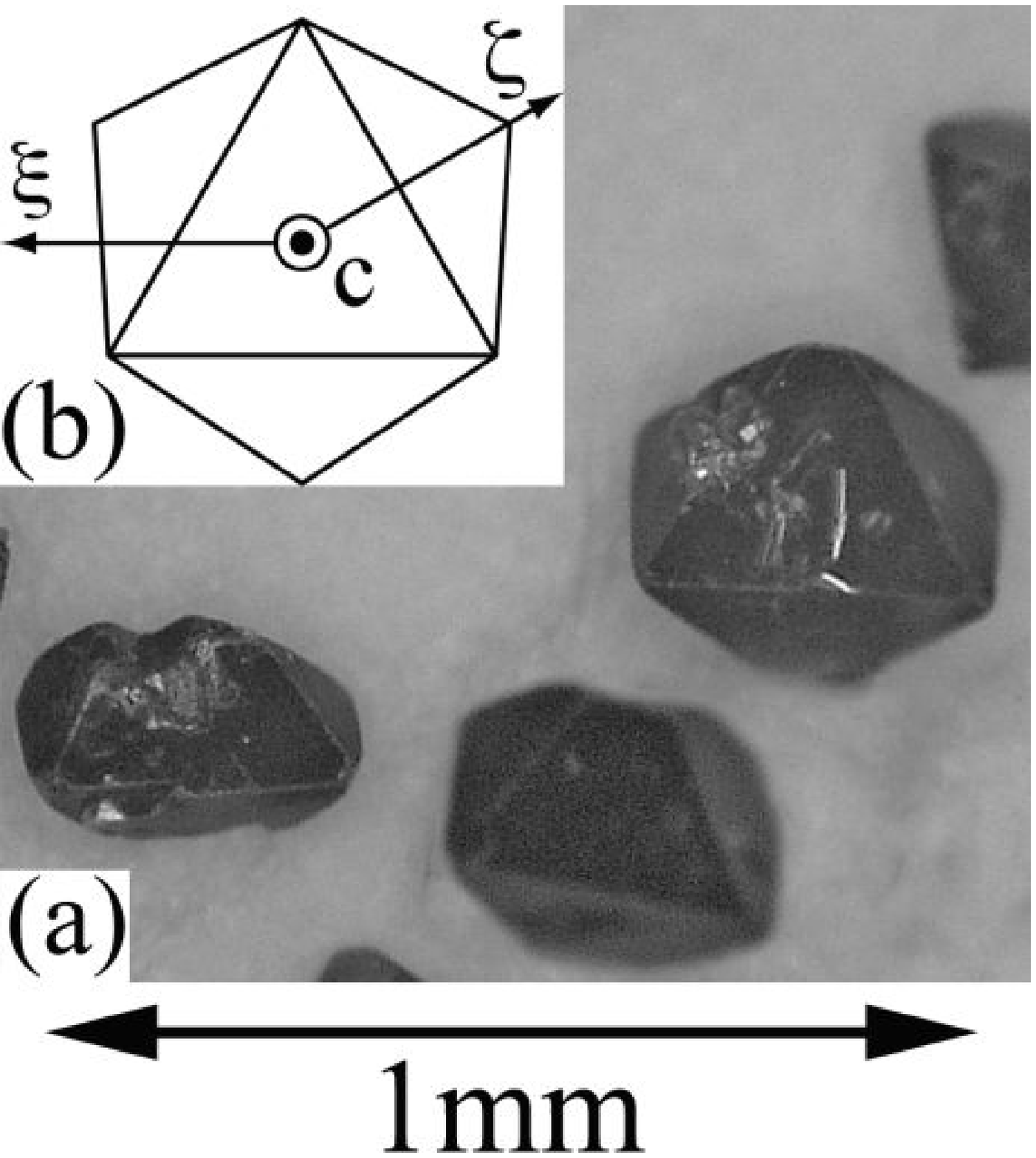}
\caption{(Color online) (a) Photograph showing the typical size of Cr-jarosite single crystals. (b) Definition of axes. The \textit{c}-axis is perpendicular to the kagome plane. 
The $\xi$-axis and the $\zeta$-axis, which are in the kagome plane, are defined as the parallel directions and perpendicular to the edge of the triangle plane of the sample shape, respectively.
}
\label{fig1}
\end{center}
\begin{center}
\includegraphics[width=7cm]{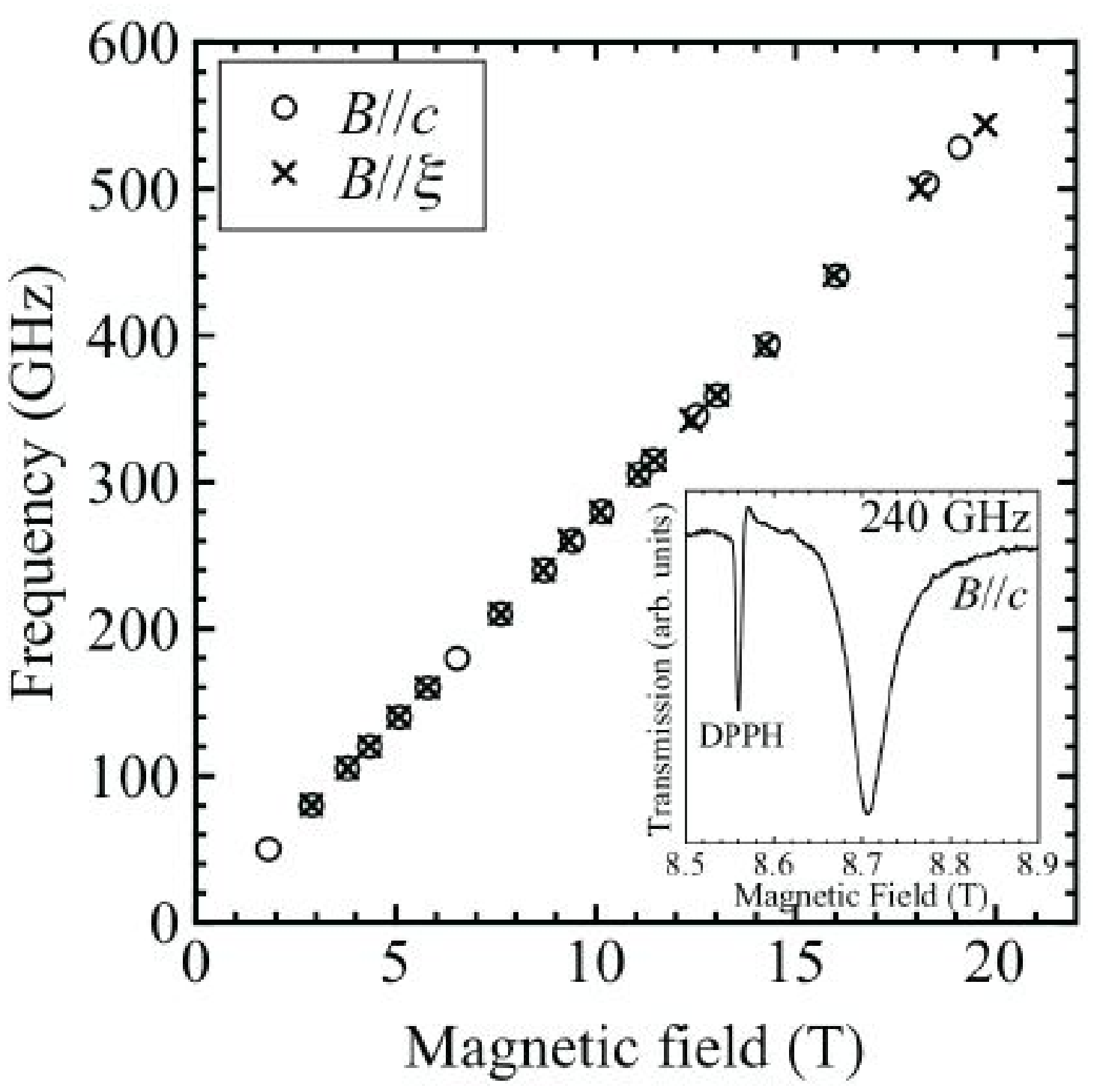}
\caption{Frequency-field diagram of Cr-jarosite for \textit{B}//\textit{c} (open circles) and \textit{B}//$\xi$ (crosses) observed at 265 K. The inset shows the typical ESR spectrum observed for \textit{B}//\textit{c} at 240 GHz. DPPH is a field marker of \textit{g}=2.}
\label{fig2}
\end{center}
\end{figure}
\indent
   On the other hand, high-frequency high-field ESR is a powerful means of studying kagome lattice model substances \cite{ref19, ref20, ref21} and determining the magnetic anisotropies of antiferromagnets \cite{ref22, ref23}. 
In particular, there are several examples \cite{ref24, ref25, ref26} in which the DM interaction is precisely determined by the analyses of high-frequency high-field ESR results. 
The aims of this study are to determine the DM interaction from high-frequency high-field ESR measurements of Cr-jarosite single crystals, and to gain a deeper insight into the ground state of Cr-jarosite.
\begin{figure}[tbp]
\begin{center}
\includegraphics[width=6cm]{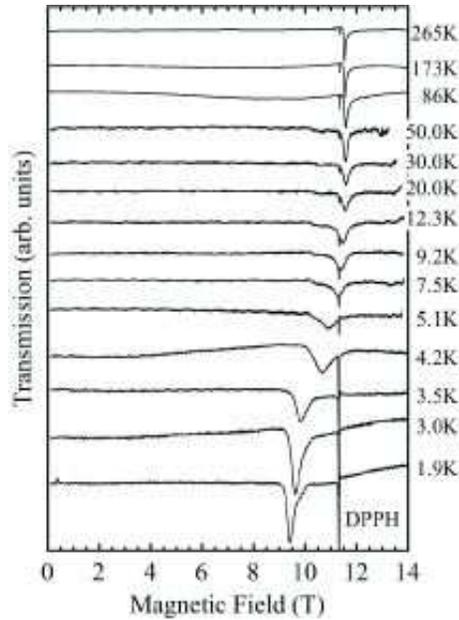}
\caption{Typical temperature dependence of ESR spectra observed for \textit{B}//\textit{c} at 317.6 GHz. DPPH is a field marker of \textit{g}=2.}
\label{fig3}
\end{center}
\end{figure}

\section{\label{sec:Experimental}Experimental}
   High-frequency high-field ESR measurements of polycrystalline and single crystals of Cr-jarosite have been performed in the temperature region from 1.9 to 265 K using a pulsed magnetic field up to 16 T.  Gunn oscillators and backward-traveling wave oscillators, which cover the frequency region from 80 to 481 GHz, have been used as light sources. Details of our high-frequency high-field ESR systems can be found in Refs. \citen{ref27, ref28, ref29, ref30}. \\
\indent
  X-band (9.5 GHz) ESR measurements have also been performed using a Bruker EMX081 ESR spectrometer in the temperature region from 5 to 296 K using a He-flow cryostat. 
Here, we should point out that the lowest temperature for X-band ESR measurement is just above the N\'eel temperature $T_\mathrm{N}$=4.5 K. 
The angular dependence measurements have been performed using a single crystal and a goniometer at 296 K. \\
\indent
   Single crystals of $\mathrm{KCr_3(OH)_6(SO_4)_2}$ (Cr-jarosite) were grown by a hydrothermal method. 
The starting materials, $\mathrm{K_2Cr_2O_7}$, $\mathrm{KClO_3}$, and $\mathrm{B_2O_3}$ were sealed in a gold capsule with 0.3 ml of sulfuric acid (2 mol/l) solution, which was kept at 723 K under a hydrostatic pressure of 150 MPa for one day. 
The crystal structure was analyzed using a cylindrical imaging-plate single-crystal X-ray diffractometer. 
As a result of the hydrothermal reaction, we obtained dark-brown crystals shaped as deformed octahedrons. 
X-ray diffraction measurement revealed a hexagonal unit cell with the lattice parameters \textit{a} = 7.226 \AA \hspace{0.001em} and \textit{c} = 17.221 \AA. 
See Fig. \ref{fig1} in Ref. \citen{ref12} for how the kagome lattice of Cr atoms is formed.
\\
\indent
   Figure \ref{fig1}(a) shows single crystals. 
The typical size is about 0.4 mm. $\xi$ and \textit{c} axes are defined as shown in Fig. \ref{fig1}(b). 
Twenty-five single crystals are aligned on a polyethylene sheet, where the \textit{c}-axis is perpendicular to the sheet, in order to increase the high-frequency high-field ESR intensity. 

\section{\label{sec:ResurltsDiscussion}Results and Discussion}
   Figure \ref{fig2} shows the frequency-field diagram observed for \textit{B}//\textit{c} and \textit{B}//$\xi$ at 265 K. 
The inset shows the typical spectrum observed for \textit{B}//\textit{c} at 240 GHz. 
As it yields relatively sharp ESR, the resonance field can be determined rather precisely. 
Therefore, we can obtain rather precise \textit{g}-values: $g_c = 1.9704 \pm 0.0002$ and $g_\xi =1.9720 \pm 0.0003$. 
These values are very typical for $\mathrm{Cr^{3+}}$ ions in the octahedral crystal field \cite{ref31}. These \textit{g}-values are consistent with X-band ESR measurements at 294 K; the results and a detailed discussion will appear in a later section.\\
\indent
   Figure \ref{fig3} shows the typical temperature dependence observed at 317.4 GHz for \textit{B}//\textit{c}. 
For simplicity, some of the observed temperatures are not shown. 
The change in the S/N ratio above 86 K and below 50 K is due to the use of different cryostats \cite{ref27, ref29} and is not intrinsic. 
The shift in resonance starts at slightly higher than the $\mathrm{N\acute{e}el}$ temperature $T_\mathrm{N}$=4.5 K. 
The shift in the resonance field clearly suggests the appearance of an internal field. The divergence of the linewidth, which is typical of antiferromagnetic order, is also observed at around $T_\mathrm{N}$.\\
\begin{figure}[tbp]
\begin{center}
\includegraphics[width=6.5cm]{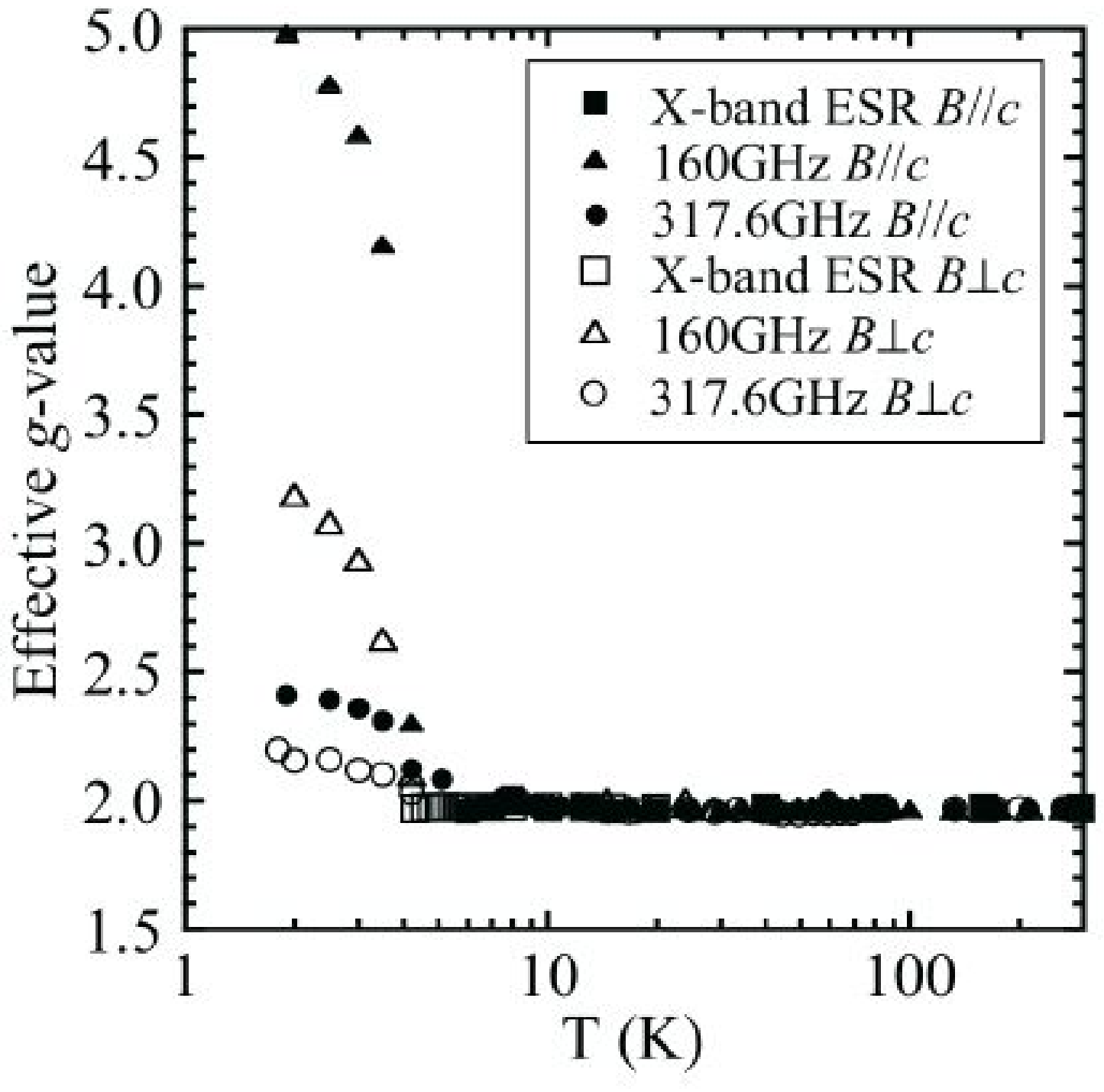}
\caption{Temperature dependences of effective \textit{g}-value.}
\label{fig4}
\end{center}
\begin{center}
\includegraphics[width=7cm]{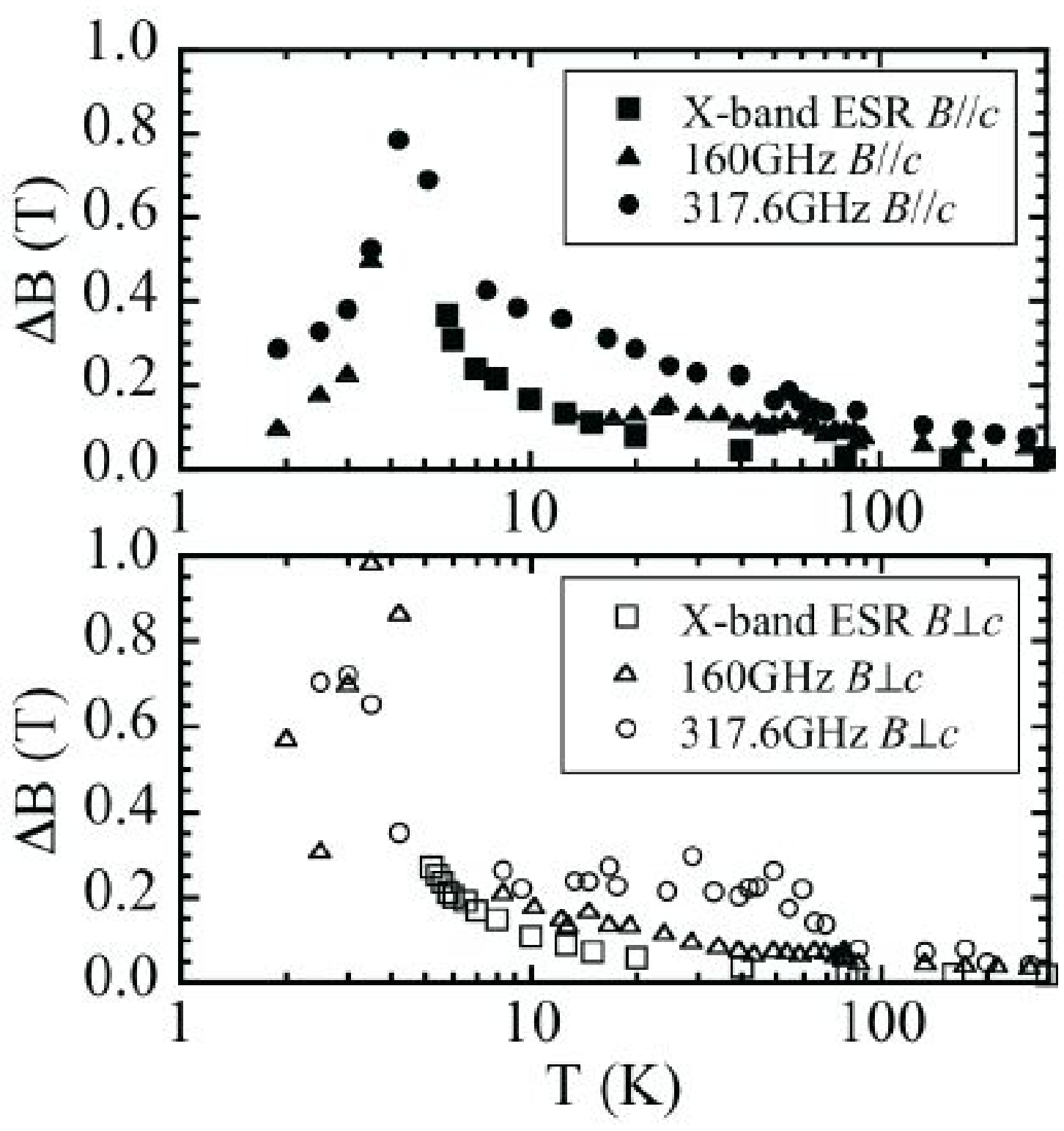}
\caption{Temperature dependences of linewidth.}
\label{fig5}
\end{center}
\end{figure}
\indent
   Figure \ref{fig4} shows the temperature dependences of the effective \textit{g}-value observed at the X-band (9.5 GHz), 160, and 317.6 GHz for \textit{B}//\textit{c} and \textit{B}$\perp$\textit{c}. 
Although the X-band measurement cannot reach temperatures below $T_\mathrm{N}$, the observed absorptions for 160 and 317.4 GHz show clear increases in the \textit{g}-value below $T_\mathrm{N}$, suggesting the presence of an internal field. 
Moreover, we can also see clear divergence of the linewidths at around $T_\mathrm{N}$ for all frequencies for \textit{B}//\textit{c} and \textit{B}$\perp$\textit{c} in Fig. \ref{fig5}. 
These results suggest that observed absorption lines below $T_\mathrm{N}$ are antiferromagnetic resonances (AFMRs) and that the critical behavior observed at around $T_\mathrm{N}$ is typical for the antiferromagnetic order.\\
\begin{figure}[h]
\begin{center}
\includegraphics[width=4.8cm]{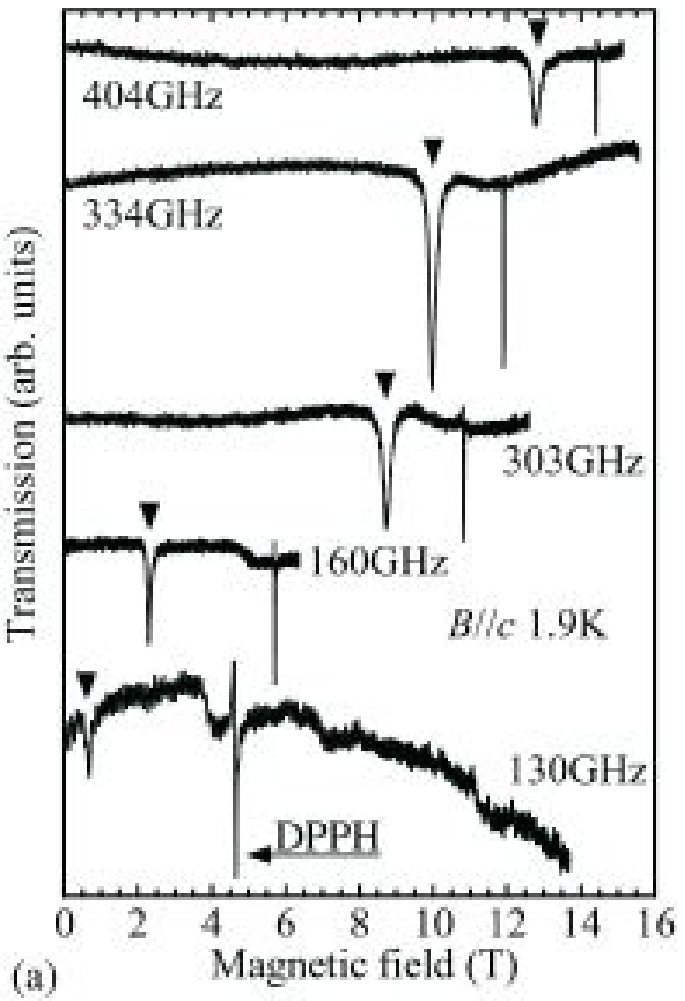}
\includegraphics[width=4.8cm]{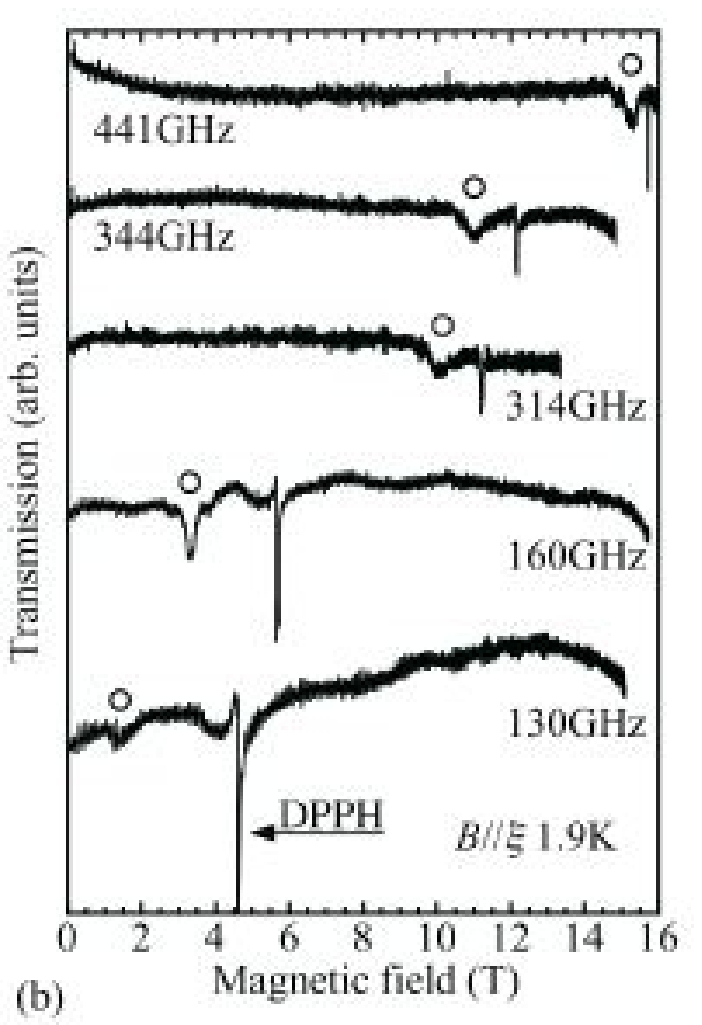}
\includegraphics[width=4.8cm]{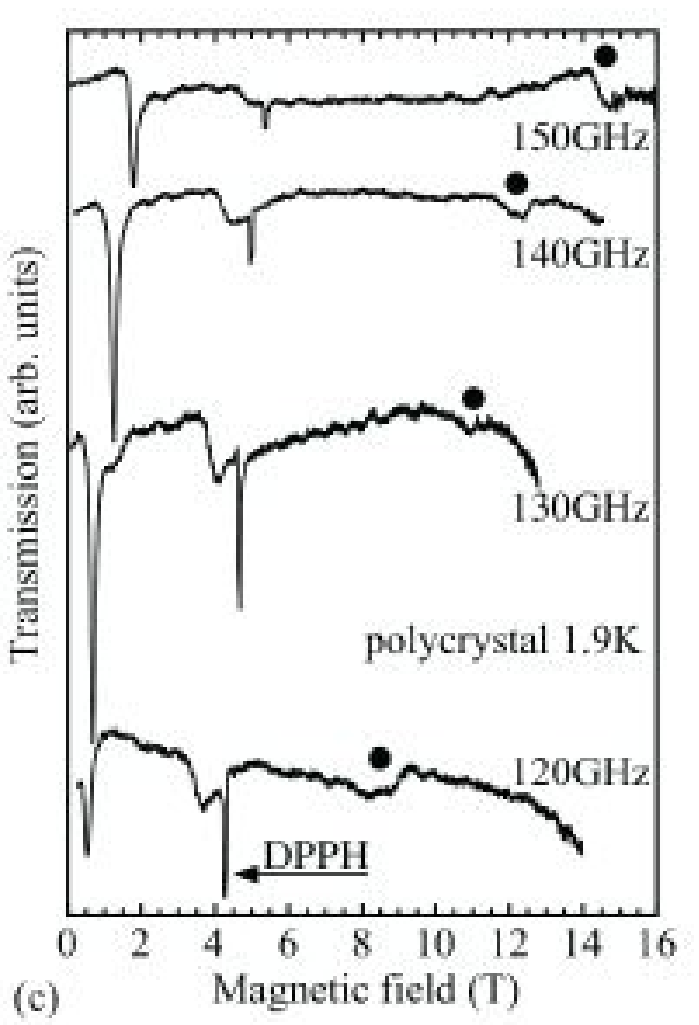}
\caption{Frequency dependence of AFMR spectra at 1.9 K for \textit{B}//\textit{c} (a), \textit{B}//$\xi$ (b), and polycrystal (c). DPPH is a field marker of \textit{g}=2.}
\label{fig6}
\end{center}
\end{figure}
\indent
   Figures \ref{fig6}(a) and \ref{fig6}(b) show typical AFMRs observed at 1.9 K for \textit{B}//\textit{c} and \textit{B}$\perp$\textit{c}. 
We can clearly see the main AFMR signals denoted by triangles and open circles. 
It is also clear that the intensities of the AFMRs are much stronger for \textit{B}//\textit{c} than for \textit{B}$\perp$\textit{c} in comparison with that of DPPH. 
The small resonances observed just below DPPH, which is a field marker of \textit{g}=2, for 130 or 160 GHz in Fig. \ref{fig6}(a) seem to be the impurity resonances because their \textit{g}-values are close to 2 even in the antiferromagnetic order phase. 
When we increased the quantity of the sample by using polycrystalline samples, we started to see weak signals, denoted by solid circles, as shown in Fig. \ref{fig6}(c). 
As these signals are observed at 1.9 K and the resonance fields are far from DPPH, these weak resonances can be considered as AFMRs. \\
\indent
   The AFMRs shown in Fig. \ref{fig6} are plotted in the frequency-field diagram of Fig. \ref{fig7}. 
The observed antiferromagnetic gap is about 120 GHz. The AFMR theory for a kagome lattice antiferromagnet on the basis of molecular field theory was discussed by Fujita \textit{et al}. \cite{ref25}, 
who considered both the DM model and the crystal field model. 
However, as it is difficult to interpret both the magnetization and AFMR results simultaneously using the crystal field model for Fe-jarosite, they used the DM model.
Because of the symmetry of the kagome lattice, DM interaction exists \cite{ref18}, and the DM vector $d_{ij}$ exists in the mirror plane between the nearest \textit{i} and \textit{j} sites in the kagome lattice, following the Moriya rule \cite{ref32}. 
Fujita \textit{et al}. considered not only the nearest-neighbor exchange interaction \textit{J} in the kagome plane and the DM interaction (see Fig. 4 in Ref. \citen{ref25} for the definition of the DM vector) but also the interplane exchange interaction $J_\perp$, and adopted six sublattices in order to explain the magnetization jump at 16.4 T observed in K-Fe jarosite. 
However, as no such magnetization jump is observed in Cr-jarosite \cite{ref12}, we will deal with the following Hamiltonian without the interplane exchange interaction and consider only three sublattices:
\begin{equation}
F = A \bm{M}_i \cdot \bm{M}_j + \bm{d} t_{ij} \cdot \bm{M}_i \times \bm{M}_j - \bm{M}_i \cdot \bm{B},
\label{eq1}
\end{equation}
where $A=6J/N(g \mu_\mathrm{B})^2$, $\bm{d} t_{ij}=6\bm{d}_{ij}/N(g\mu_\mathrm{B})^2$, and $\bm{M}_i = N g\mu_\mathrm{B} \bm{S}/3$. 
Here, \textit{N} is the number of magnetic ions and $\bm{M}_i$ is the magnetic moment on the $i$-th sublattice. 
Following the AFMR theory for \textit{B}//\textit{c} described in Ref. \citen{ref25}, %
we are able to interpret the observed AFMRs, as shown by the solid line in Fig. \ref{fig7}. 
Here, we assumed that weak AFMRs observed in Fig. \ref{fig6}(c) are AFMRs for \textit{B}//\textit{c}. 
We used the \textit{g}-value of 1.97, which was obtained at 265 K (Fig. \ref{fig2}), and \textit{J}=6.15 K, which was obtained by Okuta \textit{et al}. from the analysis of magnetic susceptibility \cite{ref12}. 
Estimated parameters are $d_\mathrm{p}=0.27$ K and $d_\mathrm{z}=0.07$ K to interpret observed the AFMRs, where $d_\mathrm{p}$ and $d_\mathrm{z}$ are the in-plane and out-of-plane components of the DM \textit{d} vector, respectively. 
The order estimation of \textit{d} can be accomplished using the following equation \cite{ref32}:
\begin{equation}
d_\mathrm{est} = \frac{ \Delta g }{ g } \, 2J = 0.09 \, \mathrm{K},
\label{eq2}
\end{equation}
where $\Delta g$ is obtained from the ESR result at 265 K. 
$d_\mathrm{est}$ is smaller than that obtained by the analysis of AFMR, but is rather acceptable for the order estimation. \\
\indent
   The existence of DM interaction causes spin canting, and spontaneous magnetization is expected at low temperatures. 
Elhajal \textit{et al}. \cite{ref18} expressed the canting angle $\eta$ [Eq. (\ref{eq3})] from the kagome plane to the \textit{c}-axis in the kagome system with competition between the exchange and DM interactions as
\begin{equation}
\tan \left( 2\eta \right) = \frac{ 2 d_\mathrm{p} }{ \sqrt{3} J + d_\mathrm{z} }.
\label{eq3}
\end{equation}
Using the parameters \textit{J}=6.15 K, $d_\mathrm{p}=0.27$ K, and $d_\mathrm{z}=0.07$ K, the obtained canting angle is $\eta$ =$1.44^\circ$. 
Using this angle, the expected spontaneous magnetization is calculated to be 0.074 $\mu_\mathrm{B}$/Cr, which is comparable to the spontaneous magnetization of 0.05 $\mu_\mathrm{B}$/Cr observed along the \textit{c}-axis obtained by Okuta \textit{et al}. \cite{ref12}. 
Therefore, we can conclude that the values of $d_\mathrm{p}$ and $d_\mathrm{z}$ obtained by the analysis of AFMR are in reasonable agreement with the results of macroscopic measurement.\\
\begin{figure}[tbp]
\begin{center}
\includegraphics[width=7cm]{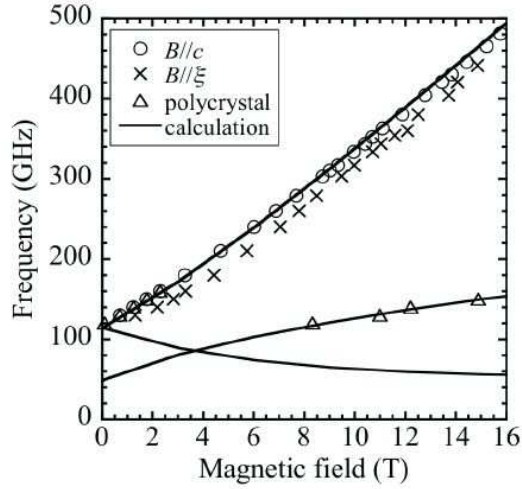}
\caption{Frequency-field diagram of AFMR at 1.9 K. Open circles, crosses, and open triangles correspond to \textit{B}//\textit{c}, \textit{B}//$\xi$, and polycrystal, respectively. The solid lines show the AFMR theory considering the DM interaction for \textit{B}//\textit{c}.}
\label{fig7}
\end{center}
\end{figure}
\begin{figure}[tbp]
\begin{center}
\includegraphics[width=6.5cm]{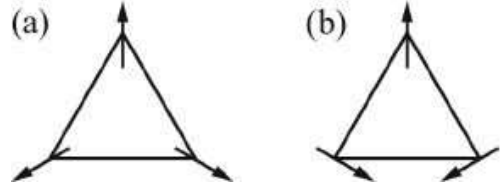}
\caption{Two possible chiralities: (a) weak ferromagnetism perpendicular to the kagome plane and (b) coplanar structure 
depending on $d_p$ and $d_z$ discussed by Elhajal \textit{et al.} 
\cite{ref18}. 
Our results suggest (a) for Cr-jarosite. See text for details. 
}
\label{fig8}
\end{center}
\end{figure}
\begin{figure}[tbp]
\begin{center}
\includegraphics[width=7cm]{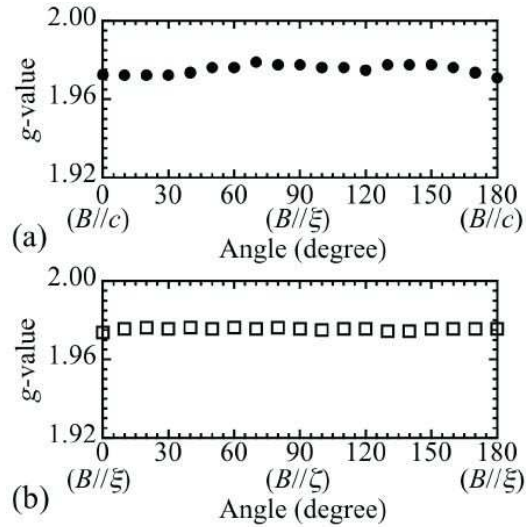}
\caption{Angular dependence of \textit{g}-values observed at 296 K using the X-band ESR. (a) Closed circles correspond to \textit{B}//\textit{c} ($0^\circ$) to \textit{B}//$\xi$ ($90^\circ$) in the \textit{c}-$\xi$ plane. (b) Open squares correspond to \textit{B}//$\xi$ ($0^\circ$) to \textit{B}//$\zeta$ ($90^\circ$) in the $\xi$-$\zeta$ plane.}
\label{fig9}
\end{center}
\end{figure}
\indent
   Next, we discuss the ground state of Cr-jarosite. 
If we compare our obtained parameters $d_\mathrm{z}/J=0.011$ and $d_\mathrm{p}/J=0.044$ with the result of the Monte Carlo simulations with classical Heisenberg spins on the kagome lattice obtained by Elhajal \textit{et al}. (Fig. \ref{fig5} in Ref. \citen{ref18}), the result clearly suggests that the ground state of Cr-jarosite corresponds to chirality (a) shown in Fig. \ref{fig8}. 
This is consistent with the  spontaneous magnetization along the \textit{c}-axis (perpendicular to the kagome plane) observed by Okuta \textit{et al}. below $T_\mathrm{N}$ \cite{ref12}, because chirality (a) possesses a weak ferromagnetism as indicated by the result obtained by Elhajal \textit{et al} \cite{ref18}. 
It is also consistent with the ground state discussed for Cr-jarosite by Wills \cite{ref18p}. 
Moreover, chirality (a) is also consistent with the magnetic structure suggested from the neutron results obtained by Inami \textit{et al}. \cite{ref15}. \\
\indent
   Finally we discuss the X-band ESR results observed at 296 K. 
As the linewidth for Cr-jarosite is relatively sharp, about 0.02 T, X-band ESR measurement using only a single crystal is possible and will enable us to obtain precise information on the anisotropy and the dimensionality of Cr-jarosite in the paramagnetic state. 
The angular dependence of the \textit{g}-value is very isotropic in the $\xi$-$\zeta$ plane (kagome plane), while small anisotropy is observed in the \textit{c}-$\xi$ plane, as shown in Fig. \ref{fig9}. 
The obtained \textit{g}-values are consistent with those obtained by high-frequency high-field ESR shown in Fig. \ref{fig2}. 
The angular dependence of the linewidth is also very isotropic in the $\xi$-$\zeta$ plane, while a very peculiar angular dependence is observed in the \textit{c}-$\xi$ plane with the minimum at around $60^\circ$ from the \textit{c}-axis, as shown in Fig. \ref{fig10}. 
Figure \ref{fig11} shows the lineshape analysis of X-band ESR spectra observed at 296 K for \textit{B}//\textit{c}, \textit{B}//$\xi$, and $60^\circ$ from \textit{B}//\textit{c} in the \textit{c}-$\xi$ plane. 
They clearly show that the lineshape is nearly Lorentzian for $60^\circ$ from \textit{B}//\textit{c} in the \textit{c}-$\xi$ plane ($c$). In the other two cases, the lineshapes are between the Gaussian and Lorentzian. \\
\indent
   Richards and Salamon discussed the ESR of a two-dimensional Heisenberg antiferromagnet ($\mathrm{K_2MnF_4}$) theoretically, assuming diffusive motion for the long-time dependence of the time correlation functions \cite{ref33}. 
They showed that the angular dependence of the ESR linewidth roughly had the form $(3 \cos^2 \theta -1)^2 + (const.)$ (see Fig. \ref{fig5} in Ref. \citen{ref33} where $\theta$ is the angle of the DC magnetic field with respect to the perpendicular to the two-dimensional plane), which is consistent with the result of the experiment with $\mathrm{K_2MnF_4}$, where the linewidth is larger at $\theta=0^\circ$ than at $\theta=90^\circ$, with a minimum at around $\theta=55^\circ$. 
This result is resonably consistent with our result in the \textit{c}-$\xi$ plane with the minimum at around $60^\circ$ from the \textit{c}-axis, as shown in Fig. \ref{fig10}(a). 
The solid line indicates the best-fit line of $\Delta B = \alpha (3 \cos ^2 \theta -1 )^2 + \beta$  with the parameters $\alpha = 21.8$ and $\beta = 175.9$.
This suggested that the angular dependence of the linewidth cannot be explained by either the secular or nonsecular parts of the second moment, which indicates that the linewidth is larger at $\theta=90^\circ$ than at $\theta=0^\circ$ (see Fig. \ref{fig2} in Ref. \citen{ref33}), which was explicitly due to the dominance of wavevector \textit{q}=0 modes in the long-time decay of correlations in a two-dimensional system. 
The small discrepancy between the fit and the data in Fig. \ref{fig10}(a) may originate from the DM interaction, which was not considered by Richards and Salamon \cite{ref33}.
Moreover, they also showed that the room-temperature lineshapes were Lorentzian at $\theta=55^\circ$ and between the Lorenztian and Gaussian at $\theta=0^\circ$, consistent with the results of the experiment with $\mathrm{K_2MnF_4}$. 
Again, this is consistent with our results in Fig. \ref{fig11}. 
From these findings, we can conclude that Cr-jarosite is a good two-dimensional system from the standpoint of ESR at 296 K. \\
\begin{figure}[tbp]
\begin{center}
\includegraphics[width=7cm]{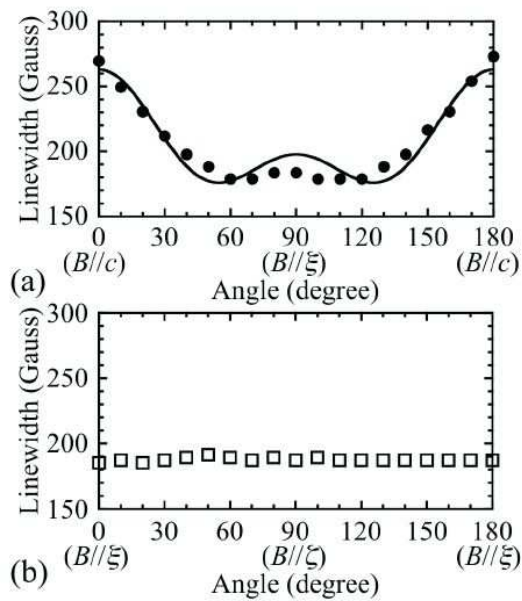}
\caption{Angular dependence of linewidths observed at 296 K using the X-band ESR. (a) Closed circles correspond to \textit{B}//\textit{c} ($0^\circ$) to \textit{B}//$\xi$ ($90^\circ$) in the \textit{c}-$\xi$ plane. 
The solid line is a fitted line.
(b) Open squares correspond to \textit{B}//$\xi$ ($0^\circ$) to \textit{B}//$\zeta$ ($90^\circ$) in the $\xi$-$\zeta$ plane.}
\label{fig10}
\end{center}
\end{figure}
\begin{figure}[tbp]
\begin{center}
\includegraphics[width=6.5cm]{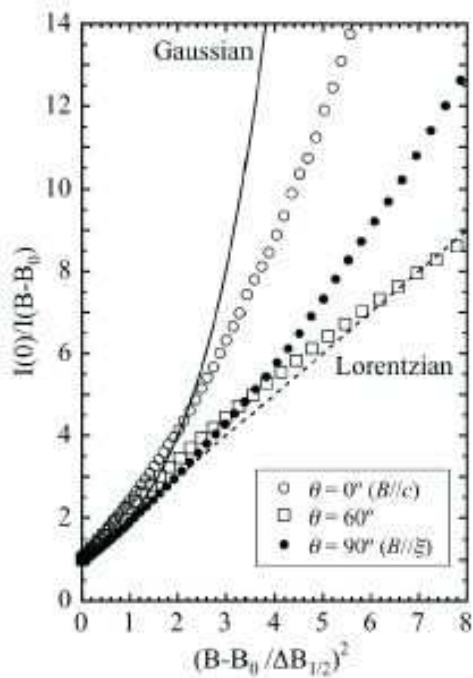}
\caption{Lineshape analysis of ESR spectra observed at 296 K using the X-band ESR. Open circles, closed circles, and open squares correspond to $\theta=0^\circ$ (\textit{B}//\textit{c}), $\theta=90^\circ$ (\textit{B}//$\xi$), and $\theta=60^\circ$ in the \textit{c}-$\xi$ plane, respectively.}
\label{fig11}
\end{center}
\end{figure}

\section{\label{sec:Conclusion}Conclusions}
   X-band and high-frequency ESR measurements of the high-quality \textit{S}=3/2 kagome lattice antiferromagnet $\mathrm{KCr_3(OH)_6(SO_4)_2}$ (Cr-jarosite) single crystals were performed for the first time. 
Taking advantage of the single crystals, we were able to determine the \textit{g}-values at 265 K as $g_c=1.9704 \pm 0.0002$ and $g_\xi=1.9720 \pm 0.0003$, the anisotropies of which are very small. 
The temperature dependence of ESR showed the divergence of the linewidth at around $T_\mathrm{N}$=4.5 K and the existence of an internal field below $T_\mathrm{N}$. 
AFMR modes with an antiferromagnetic gap of 120 GHz were revealed by high-frequency ESR measurements at 1.9 K. 
Their analysis using conventional molecular field theory yielded $d_\mathrm{p}$=0.27 K and $d_\mathrm{z}$=0.07 K, where $d_\mathrm{p}$ and $d_\mathrm{z}$ are in-plane and out-of-plane components of the DM \textit{d} vector, respectively. 
These results were discussed in connection with the Monte Carlo simulation results with classical Heisenberg spins on the kagome lattice obtained by Elhajal \textit{et al}. \cite{ref18}, and the ground state of Cr-jarosite was found to be the chirality (a) state shown in Fig. \ref{fig8} with weak ferromagnetism, which is consistent with other experimental results, including those of a neutron experiment. 
Finally, the angular dependence of the linewidth and lineshape observed at 296 K by X-band ESR were found to be very similar to those of $\mathrm{K_2MnF_4}$, which is a typical two-dimensional antiferromagnet, and to those discussed by Richards and Salamon theoretically for two-dimensional antiferromagnets. 
In conclusion, we can suggest that Cr-jarosite is an ideal model substance for an \textit{S}=3/2 Heisenberg kagome lattice antiferromagnet with good two-dimensionality from the point of view of ESR.

\begin{acknowledgments}
This work was supported by Grants-in-Aid for Scientific Research (C) No. 26400335 and No. 21550139 and for Young Scientists (B) No. 26800169 from the Japan Society for the Promotion of Science (JSPS) and by a Grants-in-Aid for Scientific Research on Priority Areas No. 19052005 "Novel States of Matter Induced by Frustration" from the Ministry of Education, Culture, Sports, Science and Technology (MEXT) of Japan. 
\end{acknowledgments}


\end{document}